%% file: IROS_2023_Jonas.tex
\documentclass[letterpaper, 10 pt, conference]{ieeeconf}  % Comment this line out if you need a4paper

\IEEEoverridecommandlockouts                              % This command is only needed if 
\usepackage{blindtext}
\usepackage[nolist]{acronym}
\usepackage[table, dvipsnames]{xcolor}
\usepackage[pdftex]{graphicx} % graphics package
\usepackage{graphicx}
\usepackage[colorinlistoftodos]{todonotes}
\usepackage{graphicx}
\usepackage{caption}
\usepackage{subcaption}

\graphicspath{{figures/}} % declare the path where graphic files are
\DeclareGraphicsExtensions{.pdf,.jpg,.png} % grafic files extensions
\usepackage{amsmath} % math package for split
\usepackage{amsfonts} % math package for fonts
\usepackage{mathtools} % math package for tools like prescript
\usepackage{tikz} % figure package
\usetikzlibrary{shapes,arrows,fit,calc,positioning,automata,backgrounds}
\usepackage{color} % colors package
\usepackage{multirow} % multiple rows in table
\usepackage{hhline} % separator in 
\usepackage{booktabs}
\usepackage[bookmarks=false, breaklinks]{hyperref} % hyperlink package
\usepackage{enumerate} % enumerate package
\usepackage{epstopdf} % package to include eps. files
\usepackage{dsfont}
\usepackage{siunitx} % package for international system units
\usepackage{cite}   % for compact citations, i.e. [1, 2]
\usepackage{diagbox}    % for making slash cell in a table
\usepackage{pdfpages}   
\usepackage{breqn}
\usepackage{svg}
\usepackage{multicol}
\svgpath{{images/svg/}} % <- using \svgpath to avoid warning
\colorlet{Green1}{green!90!}
\colorlet{Green2}{green!60!}
\colorlet{Green3}{green!40!}
\colorlet{Green4}{green!20!}
\colorlet{Green5}{green!10!}

\usepackage{url}

\usepackage{breakurl}

\RequirePackage{tikz} % For \foreach used for Orcid icon
\newcommand{\orcidicon}{\includegraphics[width=0.32cm]{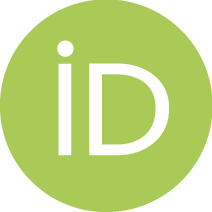}}

\foreach \x in {A, ..., Z}{%
\expandafter\xdef\csname orcid\x\endcsname{\noexpand\href{https://orcid.org/\csname orcidauthor\x\endcsname}{\noexpand\orcidicon}}
}

\definecolor{Bookcolor}{HTML}{00F9DE}
\definecolor{darkgreen}{rgb}{0.0, 0.5, 0.0}
\definecolor{gray}{gray}{0.9}
\definecolor{lightgray}{rgb}{0.86, 0.86, 0.86}

\makeatletter
\def\@citex[#1]#2{\leavevmode
\let\@citea\@empty
\@cite{\@for\@citeb:=#2\do
{\@citea\def\@citea{,\penalty\@m\ }%
\edef\@citeb{\expandafter\@firstofone\@citeb\@empty}%
\if@filesw\immediate\write\@auxout{\string\citation{\@citeb}}\fi
\@ifundefined{b@\@citeb}{\hbox{\reset@font\bfseries ?}%
\G@refundefinedtrue
\@latex@warning
{Citation `\@citeb' on page \thepage \space undefined}}%
{\@cite@ofmt{\csname b@\@citeb\endcsname}}}}{#1}}
\makeatother

\begin{acronym}
    \acro{CAMETA}{conflict-aware multi-agent estimated time of arrival}
    \acro{GNN}{graph neural network}
    \acro{ETA}{estimated time of arrival}
    \acro{AMR}{autonomous mobile robot}
    \acro{TrEPS}{traffic estimation and prediction systems}
    \acro{ITS}{intelligent transport system}
    \acro{MAPF}{multi-agent path finding}
    \acro{STSF}{spatio-temporal sequence forecasting}
    \acro{IMS}{iterated multi-step}
    \acro{DMS}{direct multi-step}
    \acro{HEAT}{heterogeneous edge-enhanced graph attention}
    \acro{GAT}{graph attention network}
    \acro{MAPE}{mean average percentage error}
    \acro{RMSE}{root mean squared error}
    \acro{MAE}{mean absolute error}
    \acro{PIBT}{priority inheritance with backtracking}
    \acro{CBS}{conflict-based search}
    \acro{HCA*}{hierarchical cooperative A*}
    \acro{WHCA*}{windowed hierarchical cooperative A*}
    \acro{LRA*}{local-repair A*}
    \acro{SOC}{sum of cost}
\end{acronym}

\begin{document}

\title{CAMETA: Conflict-Aware Multi-Agent Estimated Time of Arrival Prediction for Mobile Robots}

% use a multiple column layout for up to three different affiliations
\author{Jonas le Fevre Sejersen \orcidA{} and Erdal Kayacan \orcidB{}
% <-this % stops a space
\thanks{J. Fevre is with the Artificial Intelligence in Robotics Laboratory (AiR Lab), Department of Electrical and Computer Engineering, Aarhus University,
        8000 Aarhus C, Denmark
        {\tt\small \{jonas.le.fevre\} at ece.au.dk} E. Kayacan is with the Automatic Control Group, Department of Electrical Engineering and Information Technology, Paderborn University, Paderborn, Germany. {\tt\small \{erdal.kayacan\} at uni-paderborn.de}}%
}

\maketitle
\begin{abstract}
\input{sections/abstract}

\end{abstract}
\IEEEpeerreviewmaketitle

%\bstctlcite{IEEEexample:BSTcontrol}

% introduction and related work
\input{sections/introduction} % 1 page
\input{sections/related_work}
% 1 page
%\section{Methodology}
\input{sections/method} % 1.5 pages
\input{sections/experiments} % 0.5 pages
\input{sections/results} % 1 page
\input{sections/conclusion} % 0.25 page

%\appendix
%\section{Projection and Back-Projection}
%\input{sections/projection}

\section*{Acknowledgment}
The authors would like to acknowledge the financial and hardware contribution from Beumer Group A/S, NVIDIA Corporation, and Innovation Fund Denmark (IFD) under File No. 1044-00007B.

\bibliographystyle{IEEEtran}
\bibliography{References.bib}

\end{document}

%% file: sections/abstract.tex
% This study proposes a conflict-aware multi-agent estimate time of arrival (CAMAETA) framework for predicting the time of arrival of all agents in environments without any predefined road infrastructure. 
% The framework comprises of three layers: a traditional path planning algorithm for suggesting paths, a path evaluation layer for predicting multi-agent estimate time of arrival times for all agents, and lastly a cost layer for computing the best overall path suggested by the first layer. The novelty of the framework lies in the heterogeneous map representation along with a novel heterogeneous graph neural network architecture for multi-agent arrival time estimation. Unlike existing methods that rely on structured road infrastructure and historical data to predict time of arrival, the proposed framework utilizes total path length and basic structural features, resulting in improved generalization capabilities. To the best of our knowledge, this is the first attempt to predict multi-agent arrival times without structured road infrastructure. The results of extensive simulations demonstrate the accuracy and effectiveness of the proposed method, and show improvements in mean average percentage error of 29.5\% and 44\% when compared to a naive method that does not consider conflicts.
This study presents the conflict-aware multi-agent estimated time of arrival (CAMETA) framework, a novel approach for predicting the arrival times of multiple agents in unstructured environments without predefined road infrastructure. The CAMETA framework consists of three components: 
a path planning layer generating potential path suggestions,
a multi-agent ETA prediction layer predicting the arrival times for all agents based on the paths,
and lastly, a path selection layer that calculates the accumulated cost and selects the best path.
The novelty of the CAMETA framework lies in the heterogeneous map representation and the heterogeneous graph neural network architecture. As a result of the proposed novel structure, CAMETA improves the generalization capability compared to the state-of-the-art methods that rely on structured road infrastructure and historical data. The simulation results demonstrate the efficiency and efficacy of the multi-agent ETA prediction layer, with a mean average percentage error improvement of 29.5\% and 44\% when compared to a traditional path planning method ($A^*$) which does not consider conflicts. The performance of the CAMETA framework shows significant improvements in terms of robustness to noise and conflicts as well as determining proficient routes compared to state-of-the-art multi-agent path planners.

%% file: sections/introduction.tex
\section{Introduction}
\label{sec:introduction}
\begin{figure}[t]
   \centering
   \includegraphics[width=0.43\textwidth]{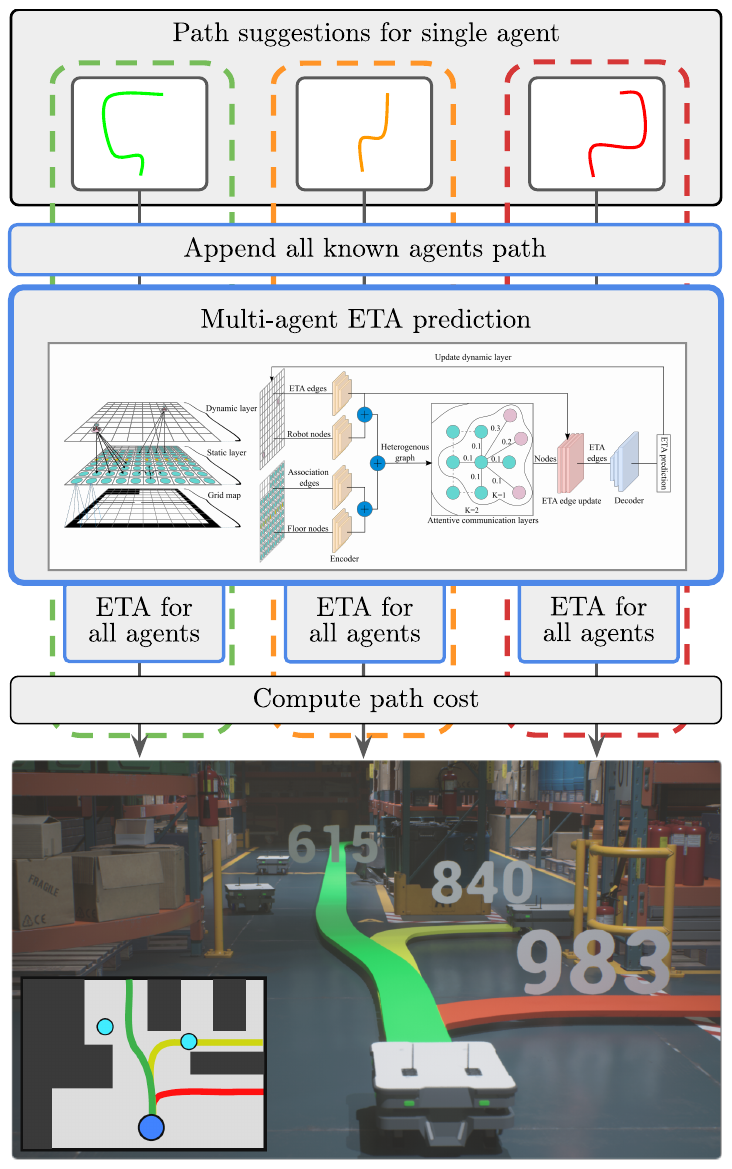}
   \caption{The conflict-aware multi-agent planner framework is depicted in three layers. The first layer generates multiple path suggestions for a single agent. The second layer, enclosed in blue borders, produces an estimated time of arrival prediction for all agents. Finally, the third layer computes the overall cost for each suggested path, using the information generated in the previous layer.}
   \label{fig:overview}
 
\end{figure}

\Ac{MAPF} is the problem of generating valid paths for multiple agents while avoiding conflicts. This problem is highly relevant in many real-world applications, such as logistics, transportation, and robotics, where multiple agents must operate in a shared environment. \ac{MAPF} is a challenging problem due to the need to find paths that avoid conflicts while minimizing the overall travel time for all agents. Many state-of-the-art \ac{MAPF} solvers \cite{CBS, PIBT, HCA*} employ various techniques to find a set of conflict-free paths on graphs representing the environment and the agents. However, a common limitation of these solvers is that they tend to generate tightly planned and coordinated paths. Therefore, the agents are expected to follow the exact path prescribed by the solver, which can lead to problems when applied to real-world systems with imperfect plan execution and uncertainties in the environment.

This work introduces a \ac{CAMETA} for indoor \ac{AMR} applications that operate in time-constrained scenarios. The proposed framework is a three-layered framework that is deployed on each agent. The layers consist of a path planning layer, which generates route suggestions for the deployed agent, a multi-agent \ac{ETA} prediction layer, which forecasts the \ac{ETA} of all agents in the system given one of the suggested paths, and a path selection layer that minimizes the overall travel time by reducing the total number of conflicts in the system. In our problem definition, some agents are required to arrive at their destination quicker than others, which is a common scenario in logistics applications for airports and warehouses. The proposed framework is illustrated in Fig. \ref{fig:overview}.

% The proposed work mainly focuses on the development of the multi-agent \ac{ETA} prediction layer and its effectiveness in forecasting the \ac{ETA} for each agent. To evaluate the multi-agent \ac{ETA} prediction layer and the proposed \ac{GNN} model, this study compares it to a naive method. In this context, the term \textit{"naive"} refers to paths, which does not consider the imperfect execution of plans and conflicts that may occur along the planned path. This comparison provides insight into the performance and accuracy gained by utilizing the \ac{GNN} model for predicting \ac{ETA} as opposed to ignoring the complexities such as imperfect execution and conflicts. Additionally, the proposed model demonstrates flexibility and generalizability by effectively adapting to different robot densities. Given the potential variations in the number of robots due to factors such as season, demand, and business requirements, the model's ability to seamlessly accommodate changes in robot densities without requiring retraining is highly valuable. This characteristic enables the industry to easily add or remove robots as needed, enhancing operational efficiency and adaptability.

The main focus of this work is on the development of the multi-agent \ac{ETA} prediction layer and its effectiveness in forecasting the \ac{ETA} for each agent. The performance and accuracy of the proposed \ac{GNN} model and the multi-agent \ac{ETA} prediction layer are evaluated by comparing them to a naive method. In this context, the term \textit{"naive"} refers to path planning that does not consider the imperfect execution of plans and conflicts that may arise along the intended path.
This comparison provides valuable insights into the performance and accuracy improvements achieved by utilizing the \ac{GNN} model for \ac{ETA} prediction, which explicitly considers the complexities associated with imperfect execution and conflicts. The comparison provides an understanding of the added value of the proposed model in accurately forecasting \ac{ETA} and addressing real-world challenges in multi-agent systems.
Furthermore, the proposed model exhibits flexibility and generalizability by effectively adapting to different robot densities. Given the potential variations in the number of robots due to seasonal fluctuations, demand changes, or business requirements, the model's ability to seamlessly accommodate shifts in robot densities without requiring retraining is of significant value. This characteristic empowers industries to easily add or remove robots as needed, thereby enhancing operational efficiency and adaptability in dynamic environments. To assess the generalizability and overall performance of the proposed framework, an experiment is conducted. This experiment aims to test how different trained prediction models scale during inference time, thereby demonstrating the framework's ability to handle varying robot densities.

In addition, an experiment is conducted to compare the overall performance of the proposed framework, including its ability to handle imperfect plan execution, with state-of-the-art \ac{MAPF} planners \cite{CBS, PIBT}. This comparative analysis serves to provide valuable insights into the effectiveness and competitiveness of the proposed framework in addressing \ac{MAPF} challenges, particularly in environments where plan execution may be imperfect. Furthermore, to ensure a comprehensive evaluation, this experiment incorporates the presence of noise, which further tests the robustness and adaptability of the proposed framework under realistic conditions. The inclusion of noise allows for a more realistic assessment of the framework's performance in the presence of uncertainties and deviations from ideal plan execution.

% Here is a small text added to destroy it all

The contributions of this study are the followings:
\begin{itemize}
    \item A conflict-aware global planner is designed to optimize overall system flow while considering time constraints for all robots in industrial scenarios.
    \item A heterogeneous graph representation is proposed to model the interaction between agents and potential bottleneck areas occurring within the map. 
    \item Finally, A novel \ac{GNN} architecture is proposed for multi-robot \ac{ETA} prediction in heterogeneous graphs supporting scaling to different robot densities during inference time. 
\end{itemize}

The rest of this paper is organized as follows. 
Section \ref{sec:problem_formulation} presents the problem formulation. Section \ref{sec:related_work} reviews the recent developments in \ac{MAPF} and \ac{ETA} prediction. 
Section \ref{sec:methodology} provides the details of the proposed framework.
Section \ref{sec:experiments} presents the experiment setup followed by the results in various grid world environments in Section \ref{sec:results}. Finally, some conclusions are drawn from this study in Section \ref{sec:conclusion}.

%% file: sections/related_work.tex
\section{Problem Formulation}
\label{sec:problem_formulation}

% MAPF definition
This section presents the formal problem formulation addressed in this paper. The problem under consideration shares similarities with the \ac{MAPF} problem \cite{Mapf}, which involves determining conflict-free paths for multiple agents on a graph to reach their respective destinations. In the \ac{MAPF} problem, time is divided into discrete timesteps, during which agents execute atomic actions synchronously, such as moving to adjacent nodes or remaining in their current locations. However, the problem addressed in this paper deviates from the traditional \ac{MAPF} problem in two key aspects. Firstly, the objective is not solely focused on finding conflict-free paths for all agents but on minimizing overall travel time by reducing total conflicts and avoiding congested areas. Secondly, a time constraint is introduced for each agent, requiring them to reach their goals before a specific deadline. This introduces the concept of priority, where higher-priority agents are assigned shorter paths.
% Type of conflicts
The types of conflicts are explained in \cite{Mapf} and describes what kind of movement patterns are not allowed to be performed and are therefore considered conflicts. The ones considered in this paper are the following four conflicts: vertex conflicts, edge conflicts, swapping conflicts, and cycle conflicts. Vertex conflicts occur when agents occupy the same position at the same time. Edge conflicts occur when agents travel along or across the same edge. Swapping conflicts occur when two agents exchange positions and cycle conflicts occur when multiple agents form a cyclic movement pattern. The notion of conflicts used in \ac{MAPF} differs from the conflicts used in the proposed work. A conflict will be defined as when an agent must alter its global path to avoid any of the four aforementioned movement patterns defined by \ac{MAPF}. 
% Environment and noise
The environment is represented as a discrete occupancy grid map, which can be viewed as a graph according to the definition in \ac{MAPF}. However, accurately modeling noise in such a setting can be challenging, as a full action would need to be performed at each time step, resulting in noise having a significant impact within a single time step. To address this challenge and simplify the incorporation of noise, the movement of the agents is modeled to operate at full velocity, and as such, the inclusion of noise in the simulation would not result in increased speed. On the contrary, the presence of noise would slow down the movement of the agents, causing them to force a wait action.
% Communication
Furthermore, it is assumed that each agent has the capability of peer-to-peer communication to resolve local conflicts using a local planner. Additionally, agents have a global connection to a state database containing each agent's committed plans and their current locations.

The formal problem formulation presented in this section sets the stage for developing effective algorithms and strategies to address the specific challenges of minimizing travel time, incorporating time constraints and priority, handling different types of conflicts, considering noise and imperfect plan execution.

\section{Related work}
\label{sec:related_work}

\subsection{Traditional \ac{MAPF} solutions}
Global path planning algorithms for the \ac{MAPF} problem are a class of solvers that first perform all computations in a single continuous interval and return a plan for the agents to follow. These plans are generated before the agents begin to move, and the agents follow the plan without any additional computation. This means that the plan cannot contain parts where agents collide. Some global solvers, such as \ac{CBS} \cite{CBS}, aim to find optimal solutions according to a predefined cost function. However, these methods may not be able to scale up to larger systems due to the exponential growth of the search space as the number of agents increases \cite{yu2013structure}. Other global algorithms, the \ac{HCA*}\cite{HCA*} and \ac{PIBT}\cite{PIBT}, sacrifice optimality or completeness in order to reduce computation time by using a spatiotemporal reservation table to coordinate agents and avoid collisions. 
A major weakness in global path planning algorithms is that agents frequently have imperfect plan execution capabilities and cannot perfectly synchronize their motions, leading to frequent and time-consuming replanning \cite{ma2017overview}. This is addressed in \cite{honig2016multi}, where a post-processing framework is proposed, using a simple temporal graph to establish a plan-execution scheduler that guarantees safe spacing between robots. The framework exploits the slack in time constraints to absorb some of the imperfect plan executions and prevent time-intensive replanning.
The work of \cite{MU} extends the method presented in \cite{M*} for a multi-agent path planner called uncertainty M* that considers collision likelihood to construct the belief space for each agent. However, this does not guarantee to remove conflicts caused by imperfect plan execution, so a local path planning algorithm is commonly needed for solving these conflicts when they occur.

Local path planning algorithms are a class of solvers that compute partial solutions in real-time, allowing agents to adjust their plans as they execute them. A simple approach for local \ac{MAPF} is \ac{LRA*} \cite{LRA*}, which plans paths for each agent while ignoring other agents. Once the agents start moving, conflicts are resolved locally by constructing detours or repairs for some agents. Another notable local \ac{MAPF} solver is the \ac{WHCA*} algorithm \cite{HCA*}, which is a local variant of the \ac{HCA*} algorithm. \ac{WHCA*} uses a spatiotemporal reservation table to coordinate agents, but only reserves limited paths, splitting the problem into smaller sections. As agents follow the partially-reserved paths, a new cycle begins from their current locations. In the traditional \ac{WHCA*}, a different ordering of agents is used in each cycle to allow a balanced distribution of the reservation table. An extension of the \ac{WHCA*} is proposed in \cite{COWHCA*}, where a priority is computed based on minimizing future conflicts, improving the success rate, and lowering the computation time. Learning-based methods for local planning are also showing promising results. In \cite{robotics11050109}, an end-to-end local planning method is introduced, using reinforcement learning to generate safe pathing in dense environments. \cite{9424371} introduces the use of \ac{GNN} for local communication and imitation learning for learning the conflict resolution of \ac{CBS} in multi-agent environments.

% \subsection{Spatio-temporal graph neural networks}
% A spatio-temporal graph combines spatio representing space or structural information and temporal representing time-varying information. 
% %Accordingly, if any system consists of a structural relationship between space and time, it can be represented as a spatio-temporal graph. These graphs can be used to design a neural network capable of handling static structures and time-varying characteristics. 

% To work with spatio-temporal graphs, we must process a sequence of graph data to build a spatio-temporal embedding that may be used for regression, classification, clustering, or correlation prediction\cite{malla2021social}. The spatial block can be any conventional \ac{GNN} \cite{battaglia2018relational}, while the temporal block can be any approach for learning over sequences of data, such as temporal convolution\cite{temporal-conv, temporal-conv-2} or temporal attention\cite{temporal-attention}. 
% Various spatio-temporal \acp{GNN}  conform to the encoding-processing-decoding paradigm \cite{GOOGLE_ETA,battaglia2018relational, keisler2022forecasting}. This paradigm accommodates the iterative nature of \ac{STSF} and pathfinding algorithms. 
% Improving these alignments has been shown to improve \acp{GNN} generalization to a more diverse distribution of graphs\cite{velivckovic2019neural}.
% 

\subsection{\ac{ETA} prediction and spatio-temporal sequence forecasting}
\label{sec:IMS-vs-DMS}
In the field of \ac{STSF}, \cite{cox1961prediction, chevillon2007direct} investigated two major learning strategies: \ac{IMS} estimation and \ac{DMS} estimation.  
The \ac{IMS} strategy trains a one-step-ahead forecasting model and iteratively uses its generated samples to produce multi-step-ahead forecasts. This strategy offers simplicity in training and flexibility in generating predictions of any length. \cite{DCRNN} demonstrated improved forecasting accuracy by incorporating graph structure into the \ac{IMS} approach.

However, the \ac{IMS} strategy suffers from the issue of accumulated forecasting errors between the training and testing phases \cite{forecast_survey}. To address this disparity, \cite{chevillon2007direct} introduced \ac{DMS} estimation directly minimizes the long-term prediction error by training distinct models for each forecasting horizon. This approach avoids error accumulation and can support multiple internal models for different horizons. Additionally, recursive application of the one-step-ahead forecaster is employed to construct multi-step-ahead forecasts, decoupling model size from the number of forecasting steps.

Although \ac{DMS} offers advantages over \ac{IMS}, it comes with increased computational complexity \cite{chevillon2007direct}. Multiple models need to be stored and trained in multi-model \ac{DMS}, while recursive \ac{DMS} requires applying the one-step-ahead forecaster for multiple steps. These factors result in greater memory storage requirements and longer training times compared to the \ac{IMS} method. On the other hand, \cite{Goodfellow-et-al-2016} shows that the \ac{IMS} training process lends itself to parallelization as each forecasting horizon can be trained independently. 

Several related works have leveraged the \ac{DMS} approach for spatio-temporal forecasting. For instance, \cite{GOOGLE_ETA} proposed a homogeneous spatio-temporal \ac{GNN} method for predicting \ac{ETA} by combining recursive \ac{DMS} and multi-model \ac{DMS}. In \cite{huang2022dueta}, a congestion-sensitive graph structure was introduced to model traffic congestion propagation, along with a route-aware graph transformer layer to capture interactions between spatially distant but correlated road segments. Furthermore, \cite{hong2020heteta} proposed a novel heterogeneous graph structure that incorporates road features, historical data, and temporal information at different scales, utilizing temporal and graph convolutions for learning spatio-temporal representations.

However, the existing methods mentioned above primarily focus on road features and consider a single vehicle traversing the graph, neglecting other types of vehicles and the influence of driver route choices on traffic conditions. Consequently, these models cannot readily be extended to multi-vehicle scenarios.

%% file: sections/method.tex
\begin{figure*}[ht]
   \centering
   \includegraphics[width=0.98\textwidth]{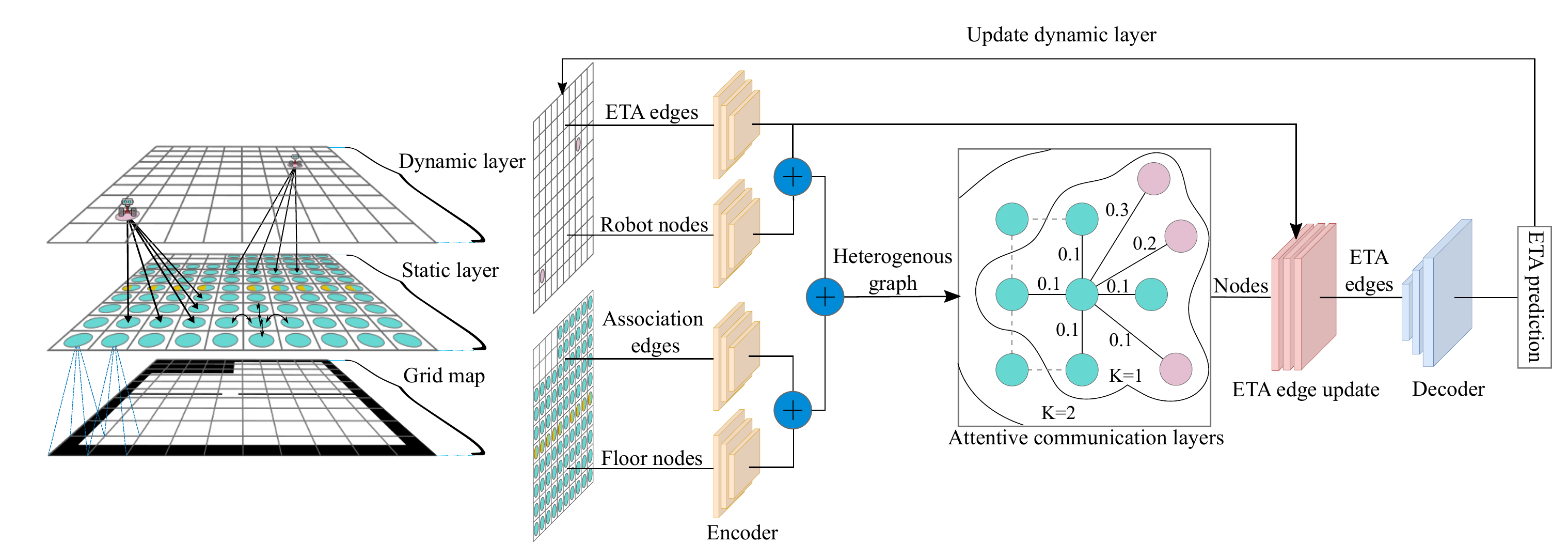}
   \caption{Illustration of the proposed multi-agent \ac{ETA} prediction layer. A simplified illustration of the three layers representing the indoor environment and the connection between the layers is shown on the left. The three layers consist of: a 2D occupancy grid map of the environment, a static layer containing all nodes and edges with static information, and a dynamic layer containing the time-variant features of the graph. The proposed \ac{GNN} architecture for multi-agent \ac{ETA} prediction is also shown.}
   \label{fig:main}
\end{figure*}
\section{Methodology}
\label{sec:methodology}
%%%%%%%%%%%%%%%%%%%%%%%%%%%%%%%%%%%%%%%%%%%%%%%%%%%%%
%%%%%%% PATHPLANNING %%%%%%%%%%%%%%%%%%%%%%%%%%%%%%%%%%%%
%%%%%%%%%%%%%%%%%%%%%%%%%%%%%%%%%%%%%%%%%%%%%%%%%%%%%
\subsection{Path planning}
\label{PP}
The path planning layer is similar in design to the \ac{LRA*}\cite{LRA*}, where other agents in the system are not considered, and resolving conflicts is left to the local planner. This allows for a decentralized global planning module, where each agent only needs to consider possible routes to its destination. While drastically improving scalability, it comes with the limitation of not being able to change the routes of other agents.
Essentially the consideration of other agents is first introduced in the multi-agent \ac{ETA} prediction layer.
% , where \ac{ETA} is computed for all agents. 
The global planner chosen for this framework is $A^*$\cite{a-star}, as it does not account for conflicts and can quickly compute different route suggestions. The proposed framework can use any off-the-shelf path-planning methods since conflicts are accounted for during the multi-agent \ac{ETA} prediction layer. It is important to note that the \ac{GNN} is trained based on the local and global path planning algorithm and, therefore, is still dependent on these. The \ac{GNN} model learns to model the interactions between these two planners, including how much the global path-planning algorithm considers conflicts in its route suggestions and how effectively the local path-planning algorithm resolves conflicts. 

It is important to note that the local planner is not directly part of the proposed pipeline but is used during execution to solve unexpected conflicts. The local planner used for training is the \ac{WHCA*} \cite{HCA*} planner with a priority-based system. The robot with the highest priority maintains its path while others adapt their paths. Each robot declares its next five moves in a local area, and local replanning is done based on priority. Agents require local peer-to-peer communication and priority-based bidding to align the reservation of space.

\subsection{Multi-agent estimated time of arrival prediction} 
The multi-agent \ac{ETA} prediction layer is the second layer of the system. The objective of the layer is to anticipate the impact of the proposed path on the other agents within the system and to estimate the additional time added by these effects. It takes as input the suggested routes from the path planning layer, as well as the planned paths of all other agents, and a 2D occupancy grid map of the environment in order to predict the \ac{ETA} for each agent. The multi-agent \ac{ETA} prediction layer is composed of two main steps: It reconstructs the data into a spatio-temporal heterogeneous graph representation and then passes the graph through a recurrent spatio-temporal \ac{GNN} model in order to make the \ac{ETA} predictions.

%%%%%%%%%%%%%%%%%%%%%%%%%%%%%%%%%%%%%%%%%%%%%%%%%%%%%%
%%%%%%% GRAPH REPRESENTATION %%%%%%%%%%%%%%%%%%%%%%%%%%%%%%%%
%%%%%%%%%%%%%%%%%%%%%%%%%%%%%%%%%%%%%%%%%%%%%%%%%%%%%%
\subsubsection{Graph representation}
% To predict multi-agent \ac{ETA} in an indoor environment, a spatio-temporal heterogeneous graph representation is proposed to capture the structure of the environment and the multiple route interaction.
To predict the \ac{ETA} of multiple agents in an indoor environment, a spatio-temporal heterogeneous graph representation is used to capture the environment's structure and the interactions between routes. The graph is divided into two layers: a static layer and a dynamic layer. The interaction between the layers is depicted in Fig. \ref{fig:main}.

The static layer of the graph only needs to be constructed once, as its features and connections remain unchanged. The static layer contains nodes of type \textit{floor} with static features of the environment, such as walls, restricted areas, and open spaces. The edges in the static layer are of type \textit{association} and represent the spatial links between nodes. The spatial characteristics of the environment are obtained from a 2D occupancy grid map, which is divided into $N\times N$ tiles. A node of type  \textit{floor} is constructed for each tile created from the given map, and the node features are the spatial information within the tile. In order to lower the number of floor nodes, all nodes containing only occupied space are removed. In the case of a solid line separating a tile, multiple \textit{floor} nodes are stacked to represent each side of the patch, as traffic on one side does not directly impact the other. These stacked nodes are depicted as multi-colored nodes in Fig. \ref{fig:main}. Association edges contain no features and are created when a direct path between two adjacent nodes or a self-loop is established.

The dynamic layer consists of nodes and edges with temporal features, where the features are subject to change or removal in the recurrent stages of the \ac{GNN} model. Nodes in this layer are of type \textit{robot}, and the temporal feature of each node is the robot's priority, computed based on the buffer time between the current time estimation and the constraint time of the robot task. If two robots have equal priority, the priority is rounded to the nearest integer and appended with the robot's internal ID as decimals to make it unique to avoid deadlocks. Edges in the dynamic layer are of type \textit{eta} and are created from a \textit{robot} node to all \textit{floor} nodes in the robot's planned trajectory. All outgoing \textit{eta} edges from a \textit{robot} node represent the planned path for the corresponding robot. The temporal features of the \textit{eta} edges include the estimated duration at each \textit{floor} node, the estimated arrival time, and the timestamp of the edge. The timestamp reflects the order in which the \ac{GNN} processes the edges. All nodes contain a self-loop edge, which has no features, to ensure that the node's own features are combined with those of its neighbors during the message-passing phase.

Combining the dynamic and static layer, the spatio-temporal heterogeneous graph is defined as follows:
\begin{equation}
G = (V, E, T_v, T_e, X_{v(t)}, X_{e(t)})  
\end{equation}
where $V$ is the set of nodes, $E$ is the set of edges, $T_v$ represents the node types belonging to the set $\{\text{robot}, \text{floor}\}$, $T_e$ represents the edge types belonging to the set $\{\text{eta},\text{association}\}$, $X_{v(t)}$ is the set of node-features at a given time $t$, and $X_{e(t)}$ is the set of edge-features at a given time $t$. 
\subsubsection{Model architecture}
The architecture of the \ac{GNN} model is created accordingly to the encode-process-decode format illustrated in Fig. \ref{fig:main}. 
First, each node and edge in the heterogeneous graph is sent to the encoding layer.
An encoder is trained for each type of node and edge in the graph to extract specific type features into a $64$-dimensional feature vector.
The encoded nodes and edges are combined into the heterogeneous graph and passed to the attentive communication layers for message passing.

In the message passing module, we utilize the \ac{HEAT} operator\cite{HEAT}.
This operator enhances the \ac{GAT} \cite{GAT} by incorporating type-specific transformations of nodes and edges features during message passing. 
Similar to \ac{GAT}, \ac{HEAT} allows running several independent attention heads to stabilize the self-attention mechanism.
The features from each attention head are concatenated as the updated output feature.
The output of the \ac{HEAT} operator is the updated node features from the neighborhood $K=1$ and the number of attention heads $H=3$ concatenated.
The updated node features are passed to the edge updater module along with the encoded \textit{eta} edges.
In the edge updater module, the \textit{eta} edges are concatenated with the corresponding nodes and sent through a linear layer.

The decoder receives the \textit{eta} edges with timestamp $T$ and converts the eta feature vectors to \ac{ETA} predictions.
For \ac{DMS} methods, the naive time of arrival of all \textit{eta} edges in the dynamic layer is updated based on the \ac{ETA} prediction to reflect the new predicted state of all robots. 
For \ac{IMS} methods, the actual arrival time is used instead of the \ac{ETA} prediction. 
% Based on either the \ac{ETA} prediction in case of \ac{DMS} or the actual arrival time in case of \ac{IMS}, the naive time of arrival of all \textit{eta} edges in the dynamic layer is updated to reflect the new state of all robots.
As this is a recurrent process, this is repeated for $T=T+1$ until all \textit{eta} edges have been decoded.

As in the first layer, the second layer does not rely on a central station to process any information. However, global information is requested from a centralized station during run-time in order to gather information about all other agents in the system. While not a completely decentralized solution, it does improve scale-ability in terms of distributing the computations to each agent.

%%%%%%%%%%%%%%%%%%%%%%%%%%%%%%%%%%%%%%%%%%%%%%%%%%%%%%%%%%%%%%%%%%%%%%%%%%
%%%%%%% COST SECTION %%%%%%%%%%%%%%%%%%%%%%%%%%%%%%%%%%%%%%%%%%%%%%%%%%%%%
%%%%%%%%%%%%%%%%%%%%%%%%%%%%%%%%%%%%%%%%%%%%%%%%%%%%%%%%%%%%%%%%%%%%%%%%%%
\subsection{Path selection and Validating time-constraints}

The path selection layer is the third layer of the proposed framework. 
This layer utilizes the \ac{ETA} to evaluate time constraints and calculate the cost of each suggested route. The layer is used to find a feasible route within a fixed time constraint and to validate the chosen route's impact on the rest of the system by utilizing the predicted \ac{ETA} for all agents. Various approaches can be used to evaluate a time-constrained path since the metric to select the best path depends on the specific application scenarios.
% and select what is considered the best path, but the choice of method depends on the specific application scenario.
In this work, a path is considered invalid if one or more of the time constraints are not met. To select between the valid paths, a cost function is proposed that considers the trade-off between the buffer times on the time constraints and the predicted arrival time. The cost function is defined as follows:

\begin{equation}
cost=\sum_{i=1}^{N}{\Bigl(\max{(\textit{TC})} - (\textit{TC}_{i} - \text{eta}_i)\Bigr)^2}
\end{equation}

\noindent where $\textit{TC}$ represents the set of all the time constraints, 
% $\textit{eta}_i$ is the i'th prediction in 
$\textit{eta}$ is the set of predicted \ac{ETA} for all $N$ robots. The cost function is designed to maximize the buffer time between the constraint time and the predicted arrival time. At the same time, it penalizes the paths with a shorter buffer time and provides more slack for the paths with a larger buffer time. This ensures that the path selection process balances the objective of meeting the time constraints with the goal of minimizing the overall time spent to complete all tasks.

%% file: sections/experiments.tex
\section{Experimental Setup}
\label{sec:experiments}
This section concisely describes the generated training and evaluation datasets and the error metrics used for evaluation and training parameters. The first set of experiments focuses on evaluating the accuracy of the \ac{ETA} prediction model, as it is a critical factor for the success of the proposed framework. The second set of experiments compares the proposed framework with state-of-the-art global planners to assess its overall performance.

\subsection{Simulation environment and noise}
% The simulation scenario is based on a discrete grid world where each agent carries out a single action at each time step. However, accurately modeling noise in such a setting can be challenging, as a full action would need to be performed at each time step, resulting in noise having a significant impact within a single time step. 
% To address this challenge and simplify the incorporation of noise, the movement of the agents is modeled to operate at full velocity. As such, the inclusion of noise in the simulation would not result in increased speed. On the contrary, the presence of noise would slow down the movement of the agents, causing them to force a wait action.
Global path planners are evaluated in terms of their robustness to conflicts and imperfections in plan execution due to noise. The experiment is designed to assess the impact of varying levels of imperfect plan execution noise, which are set to the following degrees: $\{0\%, 0.00001\%, 0.0001\%, 0.001\%, 0.01\%\}$. These levels correspond to the probability of a forced wait action resulting from accumulated noise.

We develop the \ac{ETA} prediction module without considering the presence of noise to avoid any potential bias towards a pre-defined noise level. However, during the evaluation of the overall framework, noise was introduced to all methods to conduct a more comprehensive analysis.

The experiments utilize the \ac{WHCA*} priority method as the local planner to resolve any conflicts that may arise during the simulation. This method was chosen for its simplicity, efficiency, and integration of priority selection into the algorithm. A constant time constraint is applied for all agents, prioritizing agents with longer paths. 

\subsection{Estimate time of arrival prediction}
\subsubsection{Training data}
\begin{figure}[t]
   \centering
   \includegraphics[width=0.49\textwidth]{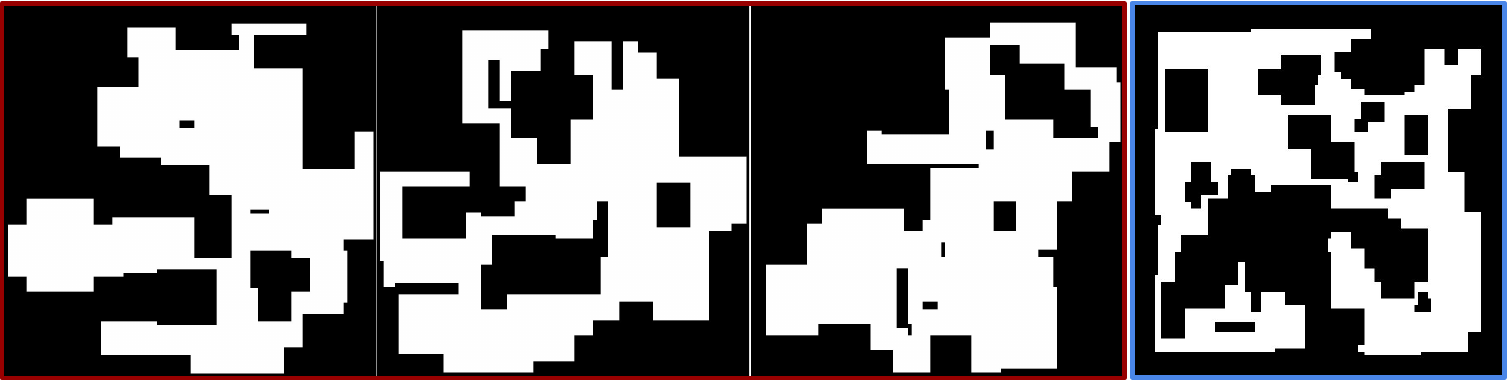}
   \caption{The figure illustrates several examples of the maps utilized in training the multi-agent \ac{ETA} prediction module (enclosed in red), and the map utilized for evaluating the overall performance of the system (enclosed in blue). The depiction of the black pixels indicates occupied spaces, while the white pixels signify unoccupied areas.}
   \label{fig:map_examples}
\end{figure}

The \ac{GNN} is trained on $5000$ unique generated warehouse environments of the size of a maximum $100\times 100$ meter. A few generated map examples are displayed in Fig. \ref{fig:map_examples}. The environments are populated with $N=\{250,500,1000\}$ robots, each with a corresponding goal point randomly distributed around the environment. 
%Different
The naive path is computed using the $A^*$ algorithm, which is used as initial input to the \ac{GNN} model. \ac{PIBT} and \ac{CBS} are not chosen as naive path planning methods for a couple of reasons. First, they modify the path of all robots and not just a single robot. Second, as the training simulation does not consider any noise or imperfect plan execution, this is the perfect scenario for these methods, resulting in no conflicts. After performing the simulation, the actual arrival time is recorded and used as the label.

\subsubsection{Evaluation criteria}
The loss function used during training is the \ac{MAPE}. \ac{MAPE} computes the percentage between predicted values $\hat{y}$ and ground truth $y$: $MAPE = \frac{1}{m}\sum_{i=1}^{m}| \hat{y_i} - y_i | / y_i.$ It is the most popular metric for \ac{ETA} prediction tasks, as the percentage penalizes the relative distance error, making it robust against outliers compared to \ac{RMSE}, which strongly penalizes the outliers with big errors.

\subsubsection{Training parameters}
The training and evaluation are conducted using an NVIDIA GeForce RTX 3090Ti GPU with 24GB VRAM. A high amount of VRAM is proven essential when training as the space required for computing the gradient of long temporal graphs is very high. However, the gradient is not computed during inference time, so a large amount of VRAM is no longer required. Due to the extensive use of VRAM, only a batch size of one is used during training. We use the Adam optimizer, while the learning rate $\gamma=0.001$ was scheduled to decay with $0.75$ every $8$th epoch. Each model has roughly trained 20 epochs, which with our hardware configuration, would take $36$ to $48$ hours.
While training takes a long time to complete, the inference time is between $400$ to $700$ ms depending on the number of robots and the length of the paths.

\subsection{Evaluation of the global path planners}
% In this experiment, the performance of global path planners is evaluated in terms of their robustness to conflicts and imperfections in plan execution due to noise. The experiment is designed to assess the impact of varying levels of imperfect plan execution noise, which are set to the following degrees: $\{0\%, 0.00001\%, 0.0001\%, 0.001\%, 0.01\%\}$. These levels correspond to the probability of a forced wait action resulting from accumulated noise.
% Four different global path planning methods will be compared for their robustness in the presence of conflicts and plan execution noise. The methods to be compared include: $A^*$, \ac{PIBT}, \ac{CBS}, and \ac{CAMETA}.

The proposed method \ac{CAMETA} is evaluated against other global path planning methods, including $A^*$, \ac{PIBT}, and \ac{CBS} with respect to their robustness in the presence of conflicts and plan execution noise.
The $A^*$ is chosen as a method that does not take into account the presence of other agents or conflicts in the paths being planned. \ac{PIBT} is chosen as a suboptimal method that is designed to be computationally efficient and considers both all agents and conflicts. \ac{CBS} is chosen as an optimal method. Although less computationally efficient than \ac{PIBT} and cannot run in real-time, it can determine the optimal paths for all agents.
% Finally, \ac{CAMETA} is a conflict-aware approach, which aims to minimize conflicts by estimating the time required to resolve the whole path, including the conflict, rather than relying on conflict-free paths.

The environment chosen for the experiment is shown in Fig. \ref{fig:map_examples} enclosed in blue. The map features both open spaces as well as corridors and potential bottlenecks. Each experiment will be conducted with $500$ agents and repeated $100$ times with different seeds of noise to eliminate a method being unlucky with the noise. However, the starting position and destination of robots will remain the same across all methods and seeds. The average \textit{makespan}, which is the average time for all robots to finish, and the \ac{SOC} are presented in the results section.

%% file: sections/results.tex
\input{LargeObjects/ETAResultTable.tex}
\section{Results}
\label{sec:results}

\subsection{Estimate time of arrival prediction}
\subsubsection{Method comparisons}
The first experiment compares the performance of the naive method, which does not use conflict-aware correction, the \ac{IMS} method, and the \ac{DMS} method. As shown in Table \ref{table:comparison}, using conflict-aware correction with either \ac{IMS} or \ac{DMS} significantly reduces error.
It can be seen that the \ac{DMS} methods perform better than \ac{IMS} methods due to the accumulated error nature in \ac{IMS},
% Comparing the \ac{IMS} and \ac{DMS} methods, it is clear that the \ac{DMS} method can learn more from the training data. It is due to the accumulated error issue in \ac{IMS},
as explained in Section \ref{sec:IMS-vs-DMS}. There is a disparity between the training and testing phases. As the density of robots rises and more conflict happens in the environment, the harder it is for the \ac{IMS} method to adapt to incorrect predictions. The best trained \ac{IMS} model improves the average \ac{MAPE} by 29.5$\%$ compared to the naive method, while the best \ac{DMS} model improves by 44$\%$. 

% The second experiment demonstrates the relationship between prediction error and path length. As depicted in Fig. \ref{fig:length-comparison},  the prediction error generated by multi-step prediction methods accumulates in proportion to the path length. Since each prediction is based on the previous prediction, the standard deviation increases rapidly as the number of prediction iterations increases. Around the path length of $175$, the error curve plateau due to less number of robots moving, and hereby less conflict occurs. The standard deviation under the \ac{MAE} curve is smaller compared to above the curve. This depicts that the majority of the predictions lie beneath the \ac{MAE} curve with a smaller error. Contrarily, the standard deviation over the \ac{MAE} curve lies farther away, indicating that the model does not accurately predict a fewer number of paths, therefore generating big standard deviation errors.

% Since the three models have roughly the same curve structure, it seems the dataset could contain some outliers that all models are having trouble predicting.
\begin{figure}[t]
   \centering
   \includegraphics[width=0.49\textwidth]{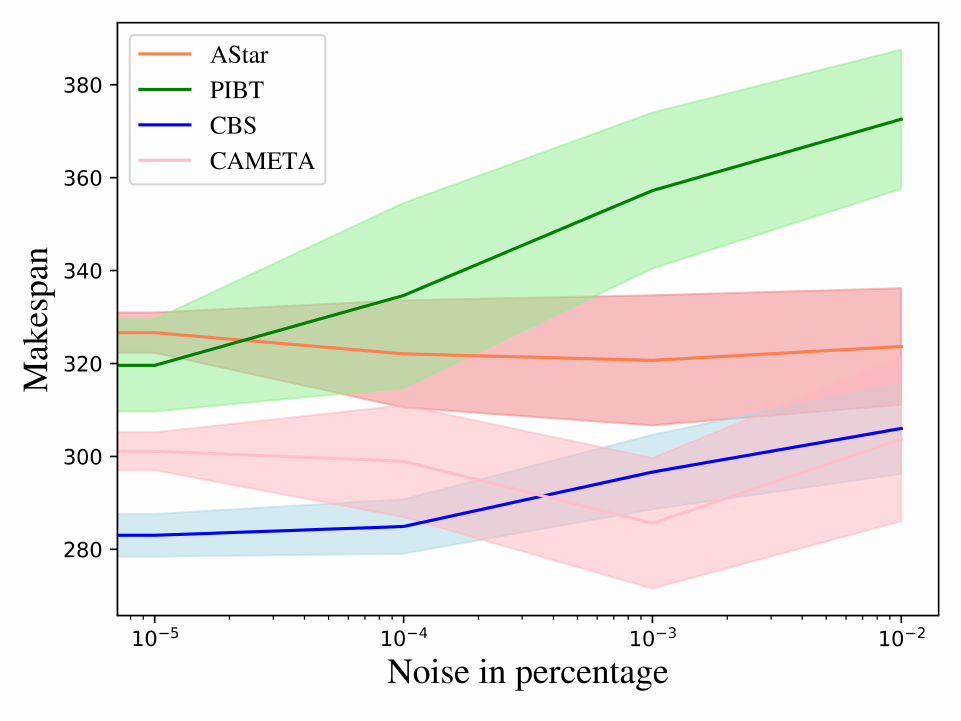}
   \caption{The plot shows the increase in the avg. makespan as noise is applied in various degrees. As the cost function prioritizes the agents with the tightest schedule, a stable makespan can be seen for \ac{CAMETA}.}
   \label{fig:noise-results-makespan}
\end{figure}
\begin{figure}[t]
   \centering\includegraphics[width=0.49\textwidth]{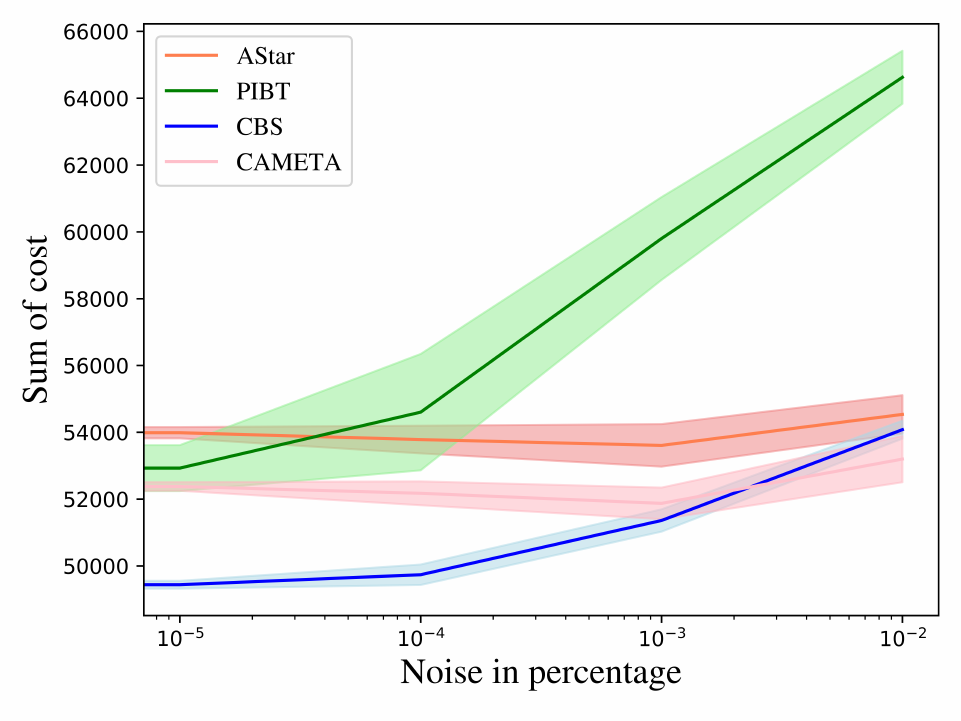}
   \caption{The plot shows the increase in avg. sum of costs as noise is applied in various degrees. Both \ac{PIBT} and \ac{CBS} rises as the noise increases, while $A^*$ and \ac{CAMETA} is less affected by the noise.}
   \label{fig:noise-results-soc}
\end{figure}
\subsubsection{Generalization under different robot densities}
The second experiment demonstrates the generalization of the models in environments of varying densities. 
% This is highly pertinent given that the number of robots can vary based on the season, demand, and business case. Therefore, for the industry to freely add or remove robots without retraining, the generalization across densities is of great importance.
As shown in Table \ref{table:comparison}, three different models trained on a dataset containing $250$, $500$, and $1000$ robots are compared at different densities. The results indicate that training in environments of high density generalizes well to environments of low density. This implies that none of the models are overfitting to any density-specific features in the graph representation. 

The \ac{DMS} models trained with the same number of robots as in the test scenario generally perform slightly better, as depicted in bold in Table \ref{table:comparison}. However, in the $500$ robot test environment, the $1000$ robot \ac{DMS} model outperforms the $500$ robot \ac{DMS} model in terms of \ac{RMSE}. As more conflicts happen in higher-density environments, the $1000$ robot \ac{DMS} model is generally trained on paths much longer than the $500$ robot \ac{DMS} model. As a result, since \ac{RMSE} is sensitive to longer paths as they accumulate more errors, as shown in experiment two, the $1000$ robot \ac{DMS} is better suited for outliers occurring within the $500$ robot test dataset.
The last column of Table \ref{table:comparison} shows average results in all the test environments. The \ac{DMS} model trained on $1000$ robots has proven to be the most generalized model.

\subsection{Evaluation of the global path planners}

The results of the path planners under the effect of noise are presented in Fig. \ref{fig:noise-results-makespan} and Fig. \ref{fig:noise-results-soc}. The \ac{CBS} method yields the optimal solution at a noise level of $0\%$, which highlights the disparity between this method and the real-time counterparts. While the \ac{CBS} algorithm demonstrates superior performance in terms of \ac{SOC} compared to other methods, its computational time is extremely long, taking four hours to generate the optimal solution.

The \ac{CAMETA} and \ac{PIBT} methods yield comparable results at $0\%$ noise, with \ac{CAMETA} outperforming \ac{PIBT} by utilizing the exploration of alternative paths with fewer conflicts. In contrast, \ac{PIBT} only resolves conflicts based on the $A^*$ algorithm, incorporating 'wait' actions and priority inheritance.
The basic $A^*$ method yields the poorest results at this noise level, as expected, due to the lack of conflict resolution in the planning. As the noise level increases, \ac{PIBT} demonstrates a decrease in performance due to the coordinated nature of its planned paths, which leaves little room for error. The additional moves planned for conflict resolution do not align with the expected conflicts, leading to even more conflicts, resulting in a rapid increase in total moves required to resolve everything as the noise level increases. This is reflected in Fig. \ref{fig:noise-results-soc} as a rapid increase in \ac{SOC} for the \ac{PIBT} method. The local planner, \ac{WHCA*}, resolves conflicts as the noise level increases. However, due to the extra moves already planned for conflicts, conflicts arising from noise are harder to resolve. Both the basic $A^*$ and \ac{CAMETA} methods demonstrate a stable pattern as the noise level increases, as they rely on conflicts being resolved in real-time, making it easier for the local planner to correct. However, \ac{CAMETA} exhibits better results in terms of total \ac{SOC} and makespan, as it reduces the number of conflicts by allowing for longer routes and distributing traffic, resulting in noise not affecting the planned path as much.

Overall, the results illustrate that while \ac{CBS} may provide the optimal solution in terms of \ac{SOC}, it is not a feasible solution for real-time applications. On the other hand, \ac{CAMETA} exhibits more promising and stable performance in terms of real-time path planning under noisy conditions.

%% file: LargeObjects/ETAResultTable.tex
\begin{table*}[!ht]
\centering
\rowcolors{3}{}{white}
\caption{Comparison of the model and prediction method.}
\label{table:comparison}
\resizebox{\textwidth}{!}{\begin{tabular}{| c | c |c c c|c c c|c c c|c c c|}
\hline
\multicolumn{2}{|c|}{Training settings} & \multicolumn{12}{c|}{Test environments with: } \\
\hline
\multirow{2}{*}{Methods}  & \multirow{2}{*}{Trained with}
& \multicolumn{3}{c|}{250 robots}
& \multicolumn{3}{c|}{500 robots} 
% & \multicolumn{3}{c|}{750 robots} 
& \multicolumn{3}{c|}{1000 robots}
& \multicolumn{3}{c|}{Average} \\ %\cline{3-18}
                                &
                                &  RMSE & MAPE[\%] & MAE 
                                &  RMSE & MAPE[\%] & MAE 
                                &  RMSE & MAPE[\%] & MAE 
                                % &  RMSE[m] & MAPE[\%] & MAE[rad]
                                &  RMSE & MAPE[\%] & MAE\\
\hline\hline
\multirow{1}{*}{Naive ($A^*$)} 
                                &        
                                & 8.06 & 8.36 & 3.59 
                                & 13.45 & 12.79 & 6.80 
                                & 42.29 & 26.08 & 24.83
                                % & NaN & NaN & NaN 
                                & 21.27 & 15.74 & 11.74\\
\hline\hline
\rowcolor{lightgray}\cellcolor{white}               
                                & \cellcolor{white} 250 robots
                                & 7.18 & 4.59 & 3.19
                                & 12.26 & 8.75 & 5.99
                                & 38.36 & 21.15 & 21.81
                                % & NaN & NaN & NaN 
                                & 19.27 & 11.50 & 10.33 \\
                                
\cellcolor{white}
                                & 500 robots
                                & 7.47 & 4.25 & 3.05
                                & 11.99 & 8.31 & 5.75
                                & 39.15 & 21.27 & 22.28
                                % & NaN & NaN & NaN 
                                & 19.54 & 11.28 & 10.36 \\
                                
                                % & 750 robots
                                % & NaN & NaN & NaN
                                % & NaN & NaN & NaN
                                % & NaN & NaN & NaN
                                % & NaN & NaN & NaN 
                                % & NaN & NaN & NaN\\
                                
\rowcolor{lightgray}\cellcolor{white}
\multirow{-3}{*}{IMS}
                                & \cellcolor{white} 1000 robots 
                                & 7.36 & 4.53 & 3.15
                                & 11.70 & 8.31 & 5.66
                                & 37.79 & 20.46 & 21.30
                                % & NaN & NaN & NaN 
                                & 18.95 & 11.1 & 10.04\\
                                
% \cellcolor{white}\multirow{-4}{*}{IMS}
%                                 & 1-1000 robots 
%                                 & NaN & NaN & NaN
%                                 & NaN & NaN & NaN
%                                 & NaN & NaN & NaN
%                                 % & NaN & NaN & NaN 
%                                 & NaN & NaN & NaN\\
\hline \hline
% \rowcolor{lightgray}
\cellcolor{white}
                                & \cellcolor{white}250 robots
                                % RMSE,  MAPE,  MAE
                                & \textbf{6.14} & \textbf{3.72} & \textbf{2.52}
                                & 9.58 & 7.09 & 4.60
                                & 31.33 & 17.17 & 17.03
                                % & NaN & NaN & NaN 
                                & 15.68 & 9.33 & 8.05 \\
                                
\rowcolor{lightgray}
\cellcolor{white}               
                                & \cellcolor{white}500 robots
                                % RMSE,  MAPE,  MAE
                                & 6.38 & 3.78 & 2.60
                                & 9.17 & \textbf{6.91} & \textbf{4.44}
                                & 28.42 & 16.16 & 15.51
                                % & NaN & NaN & NaN 
                                & 14.66 & 8.95 & 7.52\\
                                
                                % & 750 robots
                                % & NaN & NaN & NaN
                                % & NaN & NaN & NaN
                                % & NaN & NaN & NaN
                                % & NaN & NaN & NaN 
                                % & NaN & NaN & NaN\\
                                
% \rowcolor{lightgray}
\cellcolor{white} 
\multirow{-3}{*}{DMS}              
                                &\cellcolor{white} 1000 robots 
                                % RMSE,  MAPE,  MAE
                                & 6.37 & 4.03 & 2.70
                                & \textbf{9.05} & 7.02 & 4.45
                                & \textbf{25.69} & \textbf{15.39} & \textbf{13.93}
                                % & NaN & NaN & NaN 
                                & \textbf{13.70} & \textbf{8.81} & \textbf{7.03}\\
                                
% \cellcolor{white}\multirow{-4}{*}{DMS}
%                                 & 1-1000 robots 
%                                 & NaN & NaN & NaN
%                                 & NaN & NaN & NaN
%                                 & NaN & NaN & NaN
%                                 % & NaN & NaN & NaN 
%                                 & NaN & NaN & NaN\\
\hline
\end{tabular}}
\end{table*}

%% file: sections/conclusion.tex
\section{Conclusion and future work}
\label{sec:conclusion}

In this work, we develop a framework to predict conflicts and \ac{ETA} in multi-agent environments. We formulate our problem as a spatio-temporal graph focusing on edge prediction. The proposed methodology allows \ac{ETA} prediction of all robots simultaneously, which was not possible by the previously published works. Through extensive simulation experiments, the proposed method demonstrates an increase in the accuracy of the predicted arrival time. 
% While this is the first study on the system-wide estimated time of arrival, we also 
It should be noted that our proposed framework does not solve the \ac{MAPF} problem as it does not provide collision-free paths for each agent. Instead, we employ the use of \ac{ETA} in order to minimize the number of conflicts that a local path-planning algorithm needs to resolve, resulting in a more resilient method that is better equipped to handle noise.

As future work, we aim to enhance the framework by incorporating a more sophisticated robot motion model and transitioning to a continuous space simulation to better reflect real-world dynamics. Additionally, we plan to extend the framework to include multiple types of robots in a heterogeneous graph representation, incorporating dynamics, constraints, and different types of robots as nodes. Furthermore, we will incorporate the path traversed within a floor tile as a feature attribute in the \textit{eta} edges.

%% file: IROS_2023_Jonas.bbl
\begin{thebibliography}{10}
\providecommand{\url}[1]{#1}
\csname url@samestyle\endcsname
\providecommand{\newblock}{\relax}
\providecommand{\bibinfo}[2]{#2}
\providecommand{\BIBentrySTDinterwordspacing}{\spaceskip=0pt\relax}
\providecommand{\BIBentryALTinterwordstretchfactor}{4}
\providecommand{\BIBentryALTinterwordspacing}{\spaceskip=\fontdimen2\font plus
\BIBentryALTinterwordstretchfactor\fontdimen3\font minus
  \fontdimen4\font\relax}
\providecommand{\BIBforeignlanguage}[2]{{%
\expandafter\ifx\csname l@#1\endcsname\relax
\typeout{** WARNING: IEEEtran.bst: No hyphenation pattern has been}%
\typeout{** loaded for the language `#1'. Using the pattern for}%
\typeout{** the default language instead.}%
\else
\language=\csname l@#1\endcsname
\fi
#2}}
\providecommand{\BIBdecl}{\relax}
\BIBdecl

\bibitem{CBS}
G.~Sharon, R.~Stern, A.~Felner, and N.~R. Sturtevant, ``Conflict-based search
  for optimal multi-agent pathfinding,'' \emph{Artificial Intelligence}, vol.
  219, pp. 40--66, 2015.

\bibitem{PIBT}
K.~Okumura, M.~Machida, X.~D{\'e}fago, and Y.~Tamura, ``Priority inheritance
  with backtracking for iterative multi-agent path finding,'' \emph{Artificial
  Intelligence}, p. 103752, 2022.

\bibitem{HCA*}
D.~Silver, ``Cooperative pathfinding,'' in \emph{Proceedings of the aaai
  conference on artificial intelligence and interactive digital entertainment},
  vol.~1, no.~1, 2005, pp. 117--122.

\bibitem{Mapf}
R.~Stern, N.~Sturtevant, A.~Felner, S.~Koenig, H.~Ma, T.~Walker, J.~Li,
  D.~Atzmon, L.~Cohen, T.~Kumar \emph{et~al.}, ``Multi-agent pathfinding:
  Definitions, variants, and benchmarks,'' in \emph{Proceedings of the
  International Symposium on Combinatorial Search}, vol.~10, no.~1, 2019, pp.
  151--158.

\bibitem{yu2013structure}
J.~Yu and S.~LaValle, ``Structure and intractability of optimal multi-robot
  path planning on graphs,'' in \emph{Proceedings of the AAAI Conference on
  Artificial Intelligence}, vol.~27, no.~1, 2013, pp. 1443--1449.

\bibitem{ma2017overview}
H.~Ma, S.~Koenig, N.~Ayanian, L.~Cohen, W.~H{\"o}nig, T.~Kumar, T.~Uras, H.~Xu,
  C.~Tovey, and G.~Sharon, ``Overview: Generalizations of multi-agent path
  finding to real-world scenarios,'' \emph{arXiv preprint arXiv:1702.05515},
  2017.

\bibitem{honig2016multi}
W.~H{\"o}nig, T.~S. Kumar, L.~Cohen, H.~Ma, H.~Xu, N.~Ayanian, and S.~Koenig,
  ``Multi-agent path finding with kinematic constraints,'' in
  \emph{Twenty-Sixth International Conference on Automated Planning and
  Scheduling}, 2016.

\bibitem{MU}
G.~Wagner and H.~Choset, ``Path planning for multiple agents under
  uncertainty,'' in \emph{Twenty-Seventh International Conference on Automated
  Planning and Scheduling}, 2017.

\bibitem{M*}
------, ``M*: A complete multirobot path planning algorithm with performance
  bounds,'' in \emph{2011 IEEE/RSJ International Conference on Intelligent
  Robots and Systems}, 2011, pp. 3260--3267.

\bibitem{LRA*}
A.~Zelinsky, ``A mobile robot navigation exploration algorithm,'' \emph{IEEE
  Transactions of Robotics and Automation}, vol.~8, no.~6, pp. 707--717, 1992.

\bibitem{COWHCA*}
Z.~Bnaya and A.~Felner, ``Conflict-oriented windowed hierarchical cooperative
  a,'' in \emph{2014 IEEE International Conference on Robotics and Automation
  (ICRA)}.\hskip 1em plus 0.5em minus 0.4em\relax IEEE, 2014, pp. 3743--3748.

\bibitem{robotics11050109}
\BIBentryALTinterwordspacing
H.~I. Ugurlu, X.~H. Pham, and E.~Kayacan, ``Sim-to-real deep reinforcement
  learning for safe end-to-end planning of aerial robots,'' \emph{Robotics},
  vol.~11, no.~5, 2022. [Online]. Available:
  \url{https://www.mdpi.com/2218-6581/11/5/109}
\BIBentrySTDinterwordspacing

\bibitem{9424371}
Q.~Li, W.~Lin, Z.~Liu, and A.~Prorok, ``Message-aware graph attention networks
  for large-scale multi-robot path planning,'' \emph{IEEE Robotics and
  Automation Letters}, vol.~6, no.~3, pp. 5533--5540, 2021.

\bibitem{cox1961prediction}
D.~R. Cox, ``Prediction by exponentially weighted moving averages and related
  methods,'' \emph{Journal of the Royal Statistical Society: Series B
  (Methodological)}, vol.~23, no.~2, pp. 414--422, 1961.

\bibitem{chevillon2007direct}
G.~Chevillon, ``Direct multi-step estimation and forecasting,'' \emph{Journal
  of Economic Surveys}, vol.~21, no.~4, pp. 746--785, 2007.

\bibitem{DCRNN}
Y.~Li, R.~Yu, C.~Shahabi, and Y.~Liu, ``Diffusion convolutional recurrent
  neural network: Data-driven traffic forecasting,'' \emph{arXiv preprint
  arXiv:1707.01926}, 2017.

\bibitem{forecast_survey}
X.~Shi and D.-Y. Yeung, ``Machine learning for spatiotemporal sequence
  forecasting: A survey,'' \emph{arXiv preprint arXiv:1808.06865}, 2018.

\bibitem{Goodfellow-et-al-2016}
I.~Goodfellow, Y.~Bengio, and A.~Courville, \emph{Deep Learning}.\hskip 1em
  plus 0.5em minus 0.4em\relax MIT Press, 2016,
  \url{http://www.deeplearningbook.org}.

\bibitem{GOOGLE_ETA}
A.~Derrow-Pinion, J.~She, D.~Wong, O.~Lange, T.~Hester, L.~Perez, M.~Nunkesser,
  S.~Lee, X.~Guo, B.~Wiltshire \emph{et~al.}, ``Eta prediction with graph
  neural networks in google maps,'' in \emph{Proceedings of the 30th ACM
  International Conference on Information \& Knowledge Management}, 2021, pp.
  3767--3776.

\bibitem{huang2022dueta}
J.~Huang, Z.~Huang, X.~Fang, S.~Feng, X.~Chen, J.~Liu, H.~Yuan, and H.~Wang,
  ``Dueta: Traffic congestion propagation pattern modeling via efficient graph
  learning for eta prediction at baidu maps,'' \emph{arXiv preprint
  arXiv:2208.06979}, 2022.

\bibitem{hong2020heteta}
H.~Hong, Y.~Lin, X.~Yang, Z.~Li, K.~Fu, Z.~Wang, X.~Qie, and J.~Ye, ``Heteta:
  heterogeneous information network embedding for estimating time of arrival,''
  in \emph{Proceedings of the 26th ACM SIGKDD international conference on
  knowledge discovery \& data mining}, 2020, pp. 2444--2454.

\bibitem{a-star}
\BIBentryALTinterwordspacing
P.~Hart, N.~Nilsson, and B.~Raphael, ``A formal basis for the heuristic
  determination of minimum cost paths,'' \emph{{IEEE} Transactions on Systems
  Science and Cybernetics}, vol.~4, no.~2, pp. 100--107, 1968. [Online].
  Available: \url{https://doi.org/10.1109/tssc.1968.300136}
\BIBentrySTDinterwordspacing

\bibitem{HEAT}
X.~Mo, Z.~Huang, Y.~Xing, and C.~Lv, ``Multi-agent trajectory prediction with
  heterogeneous edge-enhanced graph attention network,'' \emph{IEEE
  Transactions on Intelligent Transportation Systems}, vol.~23, no.~7, pp.
  9554--9567, 2022.

\bibitem{GAT}
P.~Velickovic, G.~Cucurull, A.~Casanova, A.~Romero, P.~Lio, and Y.~Bengio,
  ``Graph attention networks,'' \emph{stat}, vol. 1050, p.~20, 2017.

\end{thebibliography}
